\documentclass{nature}
\usepackage{graphicx}
\usepackage{gensymb}
\usepackage{amsmath}
\usepackage{afterpage}
\title{Robophysical study of jumping dynamics on granular media}

\author{Jeffrey Aguilar$^{1}$ \& Daniel I.\ Goldman$^2$}

\begin{document}
\maketitle

\begin{affiliations}
 \item School of Mechanical Engineering, Georgia Institute of Technology,
 Atlanta, Georgia 30332, USA
 \item School of Physics, Georgia Institute of Technology,
 Atlanta, Georgia 30332, USA
\end{affiliations}

\begin{abstract}
Characterizing forces on deformable objects intruding into sand and soil requires understanding the solid and fluid-like responses of such substrates and their effect on the state of the object. The most detailed studies of intrusion in dry granular media have revealed that interactions of fixed-shape objects during free impact (e.g. cannonballs) and forced slow penetration can be described by hydrostatic and hydrodynamic-like forces. Here we investigate a new class of granular interactions: rapid intrusions by objects that change shape (self-deform) through passive and active means. Systematic studies of a simple spring-mass robot jumping on dry granular media reveal that jumping performance is explained by an interplay of nonlinear frictional and hydrodynamic drag as well as induced added mass (unaccounted by traditional intrusion models) characterized by a rapidly solidified region of grains accelerated by the foot. A model incorporating these dynamics reveals that added mass degrades the performance of certain self-deformations due to a shift in optimal timing during push-off. Our systematic robophysical experiment reveals both new soft matter physics and principles for robotic self-deformation and control, which together provide principles of movement in deformable terrestrial environments.
\end{abstract}


The locomotion of terrestrial animals or robots is typically studied in scenarios where unyielding environments deform the locomotor (e.g. animal\cite{alexander2003principles,blickhan1989spring} or robot\cite{Raibert.Book1986,pratt1995series,komsuoglu2014characterization,pullinz2013walking,bridge2008hopping,burdick2003minimalist} body compliance during running or jumping on hard ground), or, conversely, when non-compliant locomotors deform yielding substrates (rigid robotic hexepedal locomotion on granular media\cite{feifeipaper,li2013terradynamics}).  However, in many robotically\cite{zhang2013ground} and biologically\cite{li2012multi,moritz2005human} relevant situations like impulsive interactions during running and hopping, the deformable substrate and locomotor simultaneously affect each other's internal states.

Our previous work\cite{li2012multi,feifeipaper,li2013terradynamics,maladen2011mechanical,maladen2009undulatory} has demonstrated that dry granular media forms an excellent substrate on which to study diverse locomotor behaviors. However, even in this well-studied system, little is known about locomotor dynamics during active impulsive interactions. Many studies of fixed-shape (non-locomoting) objects impacting and penetrating dry granular media have revealed reaction forces ($F_{GM}$) that can be described by

\begin{equation}
F_{GM}=F_p(z)+\alpha v^2,
\end{equation}
where $v$ and $z$ are the object's velocity and depth, respectively\cite{Katsuragi2007,tsimring2005impact,Umbanhowar2010}. The hydrodynamic-like term, $\alpha v^2$, results from momentum transfer to the grains (significant during high speed impact\cite{euler,poncelet,robins1972new,backman1978mechanics,allen1957dynamics,forrestal1992penetration}), where $\alpha$ is the inertial drag coefficient. The hydrostatic-like force $F_p(z)$ results from frictional forces and typically scales as $k z$ for submerged or flat intruders\cite{tsimring2005impact} for slow intrusions, where $k$ characterizes the medium's penetration resistance. This hydrostatic-like term has recently been extended to a granular resistive force theory (RFT), whereby forces are predicted on objects intruding relatively slowly (where inertial effects are negligible\cite{pouliquen2009non}) with different directions and orientations\cite{li2013terradynamics}. Such work has helped explain the kinematics of slow moving locomotors\cite{li2013terradynamics,maladen2011mechanical,maladen2009undulatory}. During high speed locomotion, recent studies of free impact in dense cornstarch solutions\cite{waitukaitis2012impact} and dry granular media\cite{Katsuragi2013} as well as rapid lightweight robot running on granular media\cite{zhang2013ground} have shown the importance of hydrodynamic-like effects during high-speed interactions. One such effect includes added mass, which effectively increases the inertia of an intruder displacing material (see\cite{brennen1982review} for a review of added mass in fluids).

During such high speed movements, locomotors are often described by complex models\cite{pandy1990optimal,zajac1993muscle}. Yet even simple active-passive self-deforming objects on hard ground can exhibit rich dynamics and provide insight into more complex systems. For example, the jumping performance of a 1D actuated spring-mass hopper is sensitive to its active self-deformation strategy, which induces motion coupled to both aerial and passive spring-mass dynamics\cite{aguilar2012lift}. We therefore posit that understanding the dynamics of rapidly self-deforming objects in complex media will require new insights into both nonlinear robot dynamics and soft matter physics when inertial effects are important.


\section*{Analysis and comparison of experimental and simulated jumping}
To discover principles of impulsive granular interactions relevant to locomotion, we took a ``robophysical" approach\cite{locrobophys} by systematically varying aspects of a robot's self deformation and the substrate's properties. We constructed and measured the performance of a simple self-deforming robot consisting of a linear actuator in series with a spring performing a variety of jumping manuevers (Fig. 1a) on granular media. The simplest jumping maneuver (which we referred to as a ``single jump'' in\cite{aguilar2012lift}) is a push-off intrusion in which the motor starts at a low center of mass and forces the thrust rod down with a single-period sine-wave trajectory. On granular media, this movement induces spring compression which pushes the foot into yielding ground. The foot descends until the substrate jams, and lift-off is achieved via a single period of spring-mass oscillation (Fig. 1b).

The properties of jamming granular media depend on volume fraction, $\phi$: dry grains transition from consolidative to dilative shearing behavior within a narrow range of volume fractions ($\phi =$ 0.57 to 0.62), and their drag\cite{gravish2014force} and penetration\cite{tapia2013effect} properties vary significantly. Thus, we expected that $\phi$ would play an important role in jump height. We characterized the role of granular compaction on single jump performance by measuring jump height over a range of $\phi$ and observed a sensitive dependence of jump height to $\phi$ (Fig. 2a). Particularly, at the optimal forcing frequency, a 5 percent reduction in  $\phi$ reduced jump performance to approximately one third of the hard ground jump height. We additionally tested the role of forcing frequency, and observed a broad band of optimal frequencies (Fig. 2a inset), similar to hard ground\cite{aguilar2012lift}.

We compared experimental single jumps with a numerical model of the robot jumper in which the foot experiences granular forces, $F_{GM}$. Since little is known about the complex granular interactions of self-deforming passive/active intruders, we first applied equation (1) for $F_{GM}$ using a linear relation for $F_p(z)$. Fitting simulation to experimental jump heights with a constant $\alpha$ (see Methods, Extended Data Fig. 1a for fitting procedure) and constant $k$ yielded parameter values that were inconsistent between different jumping strategies. Also, previous experiments for slow penetration revealed that, while $F_p(z)$ was approximately linear with depth\cite{li2013terradynamics,tapia2013effect} away from boundaries\cite{stone2004stress,stone2004local}, the relationship between $F_p(z)$ and $z$ was nonlinear near the surface. Thus, we chose to empirically determine $F_p(z)$ with slow-intrusion force vs. depth measurements (Fig. 3), which revealed a nearly linear depth-dependence at low $\phi$ that became increasingly nonlinear for higher $\phi$. We modeled this as two constant penetration resistance coefficients, $k_1$ and $k_2$, where $k_1$ was the slope of a linear fit of $F_p(z)$ near the surface, and $k_2$ was the slope at deeper intrusion. Near the granular critical packing state, $\phi_c\approx 0.60$\cite{gravish2014force}, $k_1$ transitioned to a greater sensitivity to increasing $\phi$. While the values for $k_2$ showed no transition at $\phi_c$, the $k_2$ regime ($z>\delta$) exhibited an onset of force oscillations at $\phi_c$ which steadily increased with $\phi$, consistent with shearing dynamics observed in drag\cite{tapia2013effect,gravish2014force} experiments. Implementing the two-penetration resistance relation and a constant $\alpha$ in simulation was essential for agreement with experiment; using a linear $F_p(z)$ relation yielded inaccurate simulation results.

Our previous study of jumping on hard ground demonstrated that a different actuation which we called the stutter jump, consisting of a preliminary hop landing followed immediately by a push-off (Fig. 1c), outperformed the single jump while requiring lower peak actuation power\cite{aguilar2012lift}. We tested its efficacy on sand, hypothesizing that a preliminary hop would precompact the ground, increasing granular reaction forces and improving jump heights at low $\phi$. Surprisingly, this jump yielded lower heights than the single jump at low $\phi$ (Fig. 2a).

To eliminate potential transient dynamics preventing the media from relaxing into a compacted state, we introduced a delay time of $\tau=0.75$ seconds between the pull-up phase and push-off phase of the stutter jump (Fig. 1d). The delay not only improved stutter jump heights (measured with respect to initial rod height), but surpassed the single jump at low $\phi$ (Fig. 2a), suggesting that the best way to jump on loose granular media is by enhancing the single jump with a properly timed preliminary hop, locally compacting the substrate. Indeed, measuring jump heights from after the preliminary hop revealed that low $\phi$ delayed stutter jumps resembled single jumps compacted to $\phi_c$ and higher. Varying $\tau$ at low $\phi$ revealed an optimal delay time, $\tau_{opt}$, near 100 ms (Fig. 2b). This time scale represents a 5 Hz half cycle oscillation, which is near the robot's spring-mass resonance. The natural frequency is near 8 Hz on hard ground, and, during foot intrusion, is reduced at low $\phi$ to approximately 5 Hz, since the robot gains an effective one-way series spring - the $k_1$ penetration resistance of $F_p(z)$. Thus, the timing of an optimal delayed stutter jump is determined by a combination of the robot's spring-mass dynamics and the transient settling of the granular media during local compaction.

Comparing the delayed stutter experiment with the simulation revealed that the original two-penetration resistance form of $F_p(z)$ did not accurately predict jump heights. Slow intrusion force vs. depth measurements revealed that re-intrusion into previously disturbed material (even at low speeds) altered the two-penetration resistance model parameters (see Methods, Extended Data Fig. 1b). Incorporating fitted re-intrusion parameters into $F_p(z)$ produced improved simulation accuracy for the delayed stutter. However, this model did not explain the poor performance of the regular stutter jump: the simulation showed agreement at high $\phi$, but over-estimated the stutter jump heights at low $\phi$ (Fig. 2a). This deviation was particularly evident for delayed stutter jumps with $\tau<\tau_{opt}$ (Fig. 2b), suggesting unaccounted for transient granular dynamics preventing the media from relaxing into a compact state.

Thus far, we have measured $F_p(z)$ and made assumptions about the form of the hydrodynamic-like force, $\alpha v^2$, based on models in previous literature. However, we posited that a joint analysis of the granular and robot dynamics would provide insight into the mechanism that lowered the peak height of stutter jumps. We next discuss how measuring granular flow kinematics during jumping provided insight into these dynamics, which, when incorporated into our 1D jumping model, revealed the mechanism for altered jumping performance.

\section*{Evolution of a jammed granular cone}
To measure the kinematics of granular flow during jumping, we performed a particle image velocimetry (PIV) analysis on high speed videos (Supplementary Video 1) of sidewall vertical grain flow (Fig. 4a). We additionally used these PIV measurements to calculate the shear strain rate field, $\dot{\gamma}$,

\begin{equation}
\dot{\gamma} = \sqrt{\frac{1}{2}\bigg(\frac{\partial u}{\partial x}-\frac{\partial v}{\partial y}\bigg)^2+\frac{1}{2}\bigg(\frac{\partial u}{\partial y}+\frac{\partial v}{\partial x}\bigg)^2},
\end{equation}
where $u$ is horizontal velocity and $v$ is vertical velocity (Fig. 4b). We observed triangular shear bands (long boundaries of high localized shear) that were similar to other granular compression experiments and simulations\cite{le2014emergence} (see also Supplementary I and Extended Data Fig. 2). Combined with vertical grain flow (Fig. 4a) and the PIV vector field (Fig. 4b), these shear bands illustrate the dynamics of the granular media during foot intrusion. As the foot enters the media, a cone of effectively solidified grains (outlined by the shear bands) rapidly develops underneath the foot. Moving at similar downward speeds as the foot; this cone wedges surrounding material away.

Motivated by this description of the granular kinematics, we derived a geometric model of the cone's development as a flat circular intruder plows vertically into particulate media (Fig. 5a). In this model, the depth of a jammed front of grains moving with the foot grows proportionally by $\mu$ with intrusion depth, $z$. In the 1D analogy of a line of grains that collide inelastically (as introduced in\cite{waitukaitis2012impact} to describe the speed of a jamming front during rapid intrusion in a colloidal suspension), the rate, $\mu$, is inversely proportional to the separation distance between each grain relative to grain size. In dry granular media, all grains are already in contact with other grains before intrusion begins; there is no separation distance between grains. Thus $\mu$ describes a rate at which grains settle into a locally compacted solid-like state. As the foot descends and granular cone grows, the surface area of the flat portion of the cone, $A_{flat}$, decreases due to the angle, $\theta$, of the shear bands according to

\begin{equation}
A_{flat} = \pi\bigg(R^2+\bigg(\frac{\mu z}{\tan \theta}\bigg)^2-\frac{2R\mu z}{\tan \theta}\bigg),
\end{equation}
where $R$ is the foot's radius. While the shear bands fluctuated in time ($\pm 4\degree$), as observed in a previous plowing PIV experiment\cite{gravish2014force}, $\theta$ at low $\phi$ was approximately $60\degree$. 

We hypothesized that this jammed cone extended the volume of the intruder from a flat disc to a conical wedge. A resistive force theory model (RFT) proposed by Li et al.\cite{li2013terradynamics} suggests that such a change in intruder shape affects the vertical quasistatic reaction force, $F_p(z)$. Calculating RFT forces on the evolving geometry of this granular cone captured the nonlinearity in empirical measurements of $F_p(z)$ (Fig. 3a). $F_p(z)$ was calculated by summing the contributions of stress on flat surfaces, $A_{flat}$, (using the $k_1/A_{foot}$ penetration resistance) and conical surfaces, $A_{cone}$, (using $\sigma_z(60^{\circ},90^{\circ})$ from RFT\cite{li2013terradynamics}, Fig. 5b). The effective stress per unit depth for a fully developed cone, $\iint_{cone}\sigma_z(60^{\circ},90^{\circ})\delta A/A_{foot}$, coincided with $k_2/A_{foot}$ values at low and high $\phi$ (Fig. 3b). Such insights helped explain the phenomenon of rapidly diverging values of $k_1$ and $k_2$ for $\phi>\phi_c$. Flat intrusions displace grains predominantly through normal stresses, which become more difficult at higher $\phi$, where the substrate is rapidly approaching a jammed state. Above $\phi_c$, displacement through compaction is replaced by displacement through compression, and the material stiffness becomes a component of the $k_1$ penetration resistance. Once the cone forms, the intruder produces lateral grain displacements and shear stresses. As $\phi$ increases, more grain-grain frictional contacts during shearing result in an increase in $k_2$. However, for $\phi>\phi_c$, $k_2$ is not as large as $k_1$, since shear stresses do induce as much material compression as normal stresses.
%

\section*{Emergence of added mass and inertial drag from a growing granular cone}
While the characteristics of $F_p(z)$ are insufficient to explain the transient dynamics that decrease the stutter jump height, such insights into the extended intruder volume suggest that the additional mass of the granular cone, or added mass, $m_a$, must be considered in the momentum of the foot. Added mass can contribute to a shear-thickening response in dense suspensions\cite{waitukaitis2012impact}. In the realm of actively forced impacts, added mass effects contribute to the impulse developed during the slap phase of a basilisk lizard running on water\cite{glasheen1996hydrodynamic}.

Added mass for an intruder impacting a fluid has been approximated by the hemispherical volume of liquid accelerated forward in front of the intruder, consistent with the velocity change imparted by an inelastic collision with a mass equal to the added mass\cite{richardson1948impact,wagner1932phenomena}. Similarly, by dividing the granular momentum, $P_{grains}$, by the velocity of the foot, we considered added mass in the granular media to be comprised of the grains moving with flow kinematics most similar to the downward motion of the foot. Previous studies have utilized PIV to estimate the momentum of added mass in fluids \cite{sakakibara2004stereo} and qualitatively characterize momentum transfer in dense suspensions\cite{waitukaitis2012impact}. We estimated $P_{grains}$ by spatially integrating the PIV velocity field according to $P_{grains} \approx\rho\phi\int_0^H \! \int_0^{2\pi} \! \int_0^R v(r,h)r\,dr\,d\psi\,dh$ where $h$ and $r$ are the 2D velocity field coordinates, and $\rho \approx 1000$ kg/m$^3$ is the density of poppy seeds. $\psi$ was approximated by assuming azimuthal symmetry of the flow field. The foot imparted a significant amount of momentum onto the grains proportional to the foot speed, most notably during the stutter jump (Fig. 4a right inset, maroon, Supplementary Video 2). The added mass, comprised primarily by the granular cone, reached values over four times the foot mass (Fig. 6a).

Recently, Katsuragi et al. posited that added mass forces could play a role in the dynamics of non-forced impact into dry granular media\cite{Katsuragi2013} but no experimental tests were conducted. To test the role of added mass during jumping, we modified $F_{GM}$ to incorporate these dynamics into the 1D jumping simulation. Inertial drag during granular impact originates from the momentum change associated with colliding inelastically with a virtual mass\cite{poncelet}, which accumulates when the impactor accelerates surrounding material, $\frac{d(m_av)}{dt} = \frac{d m_a}{dt}v + m_aa$. Thus, our granular reaction force becomes

\begin{equation}
F_{GM} = F_p(z) - \frac{dm_a}{dt}v - m_aa,
\end{equation}
where $a$ is the foot's acceleration. We then formulated a description of added mass accumulation based on our geometric cone model (Fig. 5a), where a differential increase in intrusion depth corresponded to a differential increase in added mass according to the following relation,

\begin{equation}
\Delta m_a = \phi\rho A_{flat} \mu\Delta z,
\end{equation}
where $\phi$ and $\rho$ are the volume fraction and grain density, respectively, and $\mu\Delta z$ is the differential depth of the jamming front. Taking the infinitesimal change in $z$, we integrated equation (5) and found the added mass to be

\begin{equation}
m_a(z) = \phi\rho\mu z\pi\bigg(R^2+\frac{1}{3}\frac{(\mu z)^2}{\tan^2\theta}-\frac{R\mu z}{\tan\theta}+C\bigg).
\end{equation}

We used this equation until $\mu z = R \tan \theta$, at which point the cone was fully formed and only the constant, $C$, contributed to an increasing growth of added mass due to extra added mass from slower moving grains surrounding the cone. Both $\mu$ and $C$ were tuned to match PIV added mass measurements (Fig. 6a). While estimating added mass in fluids can be challenging for all but simple intruder shapes\cite{richardson1948impact}, we expect that the geometry and dynamics of granular jamming fronts in other intrusion scenarios will be determined by predictable shearing behavior that forms granular cones.

Similar to sphere impact in fluids\cite{richardson1948impact}, an added mass model which is dependent on depth instead of time allows us to express the conservation of momentum force, $\frac{d m_a}{dt}v$, as $\frac{d m_a}{dz}v^2$, resulting in a depth dependent inertial drag term, $\alpha(z)v^2$, where $\alpha(z) = b\frac{dm_a}{dz}$. Inertial drag was also observed to be sensitive to depth in granular sphere\cite{Umbanhowar2010} and disk impact\cite{clark2013granular}. The constant, $b$, is a scaling coefficient required to obtain agreement between simulation and experiment. We posit that this scaling (where $b>1$ for all $\phi$) is the result of the system experiencing more inelastic granular collisions than is evident from the increasing added mass, with the cone constantly gaining and shedding grains at the shearing boundaries. Nevertheless, our added mass equation dictates that $\alpha(z)$, which is proportional to the slope of $m_a(z)$, is greatest near the surface. Introducing this reactive force into the jumping model with a correctly scaled $b$ preserved the accuracy of the single and delayed stutter jumps and appropriately decreased the stutter jump heights at low $\phi$ (Fig. 6b). While added mass effects were negligible at high $\phi$, jump heights were sensitive to the scaling, $b$, of inertial drag, especially during high frequency motor forcing.  We expect such inertial effects will also help explain other high-speed movements such as running\cite{pullinz2013walking}.

\section*{Coupling of robotic spring-mass and added mass dynamics}
We now discuss the mechanism by which the above granular physics affects the locomotor's internal state to reduce jumping performance. Added mass lowers stutter jump heights by altering the phasing of the robot's spring-mass vibration (Fig. 6c-g), in which grain momentum causes the peak spring forces to occur at a non-ideal phase of the motor's oscillation. After the preliminary hop, the foot lands and stops due to granular reaction forces (Fig. 6d). The robot's actuator continues falling while it pushes the rod down, causing spring compression as the foot encounters high inertial drag due to a rapidly developing cone of added mass (Fig. 6e). The spring reaches peak compression (Fig. 6f), slowing the thrust rod, and pushing the foot down further, assisted inertially by a fully formed added mass cone. The foot descends further due to slower decelerations from added mass (Fig. 6g), and a less compressed spring now produces smaller upward propulsion forces as the robot's center of mass takes off (See Extended Data Fig. 3 for comparison to other $F_{GM}$ models).

Prescribing a delay improves the jump height in two ways. A sufficiently long delay time separates both methods of granular intrusion: passive intrusion from the robot's falling inertia during landing and active intrusion during push-off. Separating these two mechanisms reduces the overall intrusion speeds of the foot, reducing the compounding effect of the added mass decreasing the deceleration rate. As such, the robot sinks less and is able to provide more upward spring forces to the robot. Selecting the optimal delay time ensures that the phasing transfers maximal spring energy during the upward take-off movement of the motor.

%
%
%

\begin{addendum}
 \item[Supplementary Information] available online.
 \item[Acknowledgments] This work was supported by NSF Physics of Living Systems, Burroughs Wellcome Fund, and the Army Research Office. We thank A. Karsai for assistance in simulation work as well as P. Umbanhowar and L. London for insightful comments and discussion.
 \item[Author Contributions] J.A and D.I.G. conceived the study and wrote the paper. J.A. performed the experimental work, designed and ran the simulation models, and analysed the results.
 \item[Author Information] Reprints and permissions information is available at www.nature.com/reprints. The authors declare no competing interests. Correspondence and requests for materials should be addressed to J.A. (jeffrey.aguilar@gatech.edu) or D.I.G. (daniel.goldman@physics.gatech.edu).
\end{addendum}

\bibliography{main}

\begin{thebibliography}{10}
\expandafter\ifx\csname url\endcsname\relax
  \def\url#1{\texttt{#1}}\fi
\expandafter\ifx\csname urlprefix\endcsname\relax\def\urlprefix{URL }\fi
\providecommand{\bibinfo}[2]{#2}
\providecommand{\eprint}[2][]{\url{#2}}

\bibitem{alexander2003principles}
\bibinfo{author}{Alexander, R.~M.}
\newblock \emph{\bibinfo{title}{Principles of animal locomotion}}
  (\bibinfo{publisher}{Princeton University Press}, \bibinfo{year}{2003}).

\bibitem{blickhan1989spring}
\bibinfo{author}{Blickhan, R.}
\newblock \bibinfo{title}{The spring-mass model for running and hopping}.
\newblock \emph{\bibinfo{journal}{Journal of biomechanics}}
  \textbf{\bibinfo{volume}{22}}, \bibinfo{pages}{1217--1227}
  (\bibinfo{year}{1989}).

\bibitem{Raibert.Book1986}
\bibinfo{author}{Raibert, M.}
\newblock \emph{\bibinfo{title}{Legged robots that balance}}.
\newblock MIT Press series in artificial intelligence (\bibinfo{publisher}{MIT
  Press}, \bibinfo{address}{Boston}, \bibinfo{year}{1986}).

\bibitem{pratt1995series}
\bibinfo{author}{Pratt, G.~A.} \& \bibinfo{author}{Williamson, M.~M.}
\newblock \bibinfo{title}{Series elastic actuators}.
\newblock In \emph{\bibinfo{booktitle}{Intelligent Robots and Systems 95.'Human
  Robot Interaction and Cooperative Robots', Proceedings. 1995 IEEE/RSJ
  International Conference on}}, vol.~\bibinfo{volume}{1},
  \bibinfo{pages}{399--406} (\bibinfo{organization}{IEEE},
  \bibinfo{year}{1995}).

\bibitem{komsuoglu2014characterization}
\bibinfo{author}{Komsuoglu, H.}, \bibinfo{author}{Majumdar, A.},
  \bibinfo{author}{Aydin, Y.~O.} \& \bibinfo{author}{Koditschek, D.~E.}
\newblock \bibinfo{title}{Characterization of dynamic behaviors in a hexapod
  robot}.
\newblock In \emph{\bibinfo{booktitle}{Experimental Robotics}},
  \bibinfo{pages}{667--684} (\bibinfo{organization}{Springer},
  \bibinfo{year}{2014}).

\bibitem{pullinz2013walking}
\bibinfo{author}{Qian, F.} \emph{et~al.}
\newblock \bibinfo{title}{Walking and running on yielding and fluidizing
  ground}.
\newblock \emph{\bibinfo{journal}{RSS}} \bibinfo{pages}{345--353}
  (\bibinfo{year}{2013}).

\bibitem{bridge2008hopping}
\bibinfo{author}{Bridge, B.}, \bibinfo{author}{Dubowsky, S.},
  \bibinfo{author}{Kesner, S.}, \bibinfo{author}{Plante, J.-S.} \&
  \bibinfo{author}{Boston, P.}
\newblock \bibinfo{title}{Hopping mobility concept for search and rescue
  robots}.
\newblock \emph{\bibinfo{journal}{Industrial Robot: An International Journal}}
  \textbf{\bibinfo{volume}{35}}, \bibinfo{pages}{238--245}
  (\bibinfo{year}{2008}).

\bibitem{burdick2003minimalist}
\bibinfo{author}{Burdick, J.} \& \bibinfo{author}{Fiorini, P.}
\newblock \bibinfo{title}{Minimalist jumping robots for celestial exploration}.
\newblock \emph{\bibinfo{journal}{The International Journal of Robotics
  Research}} \textbf{\bibinfo{volume}{22}}, \bibinfo{pages}{653--674}
  (\bibinfo{year}{2003}).

\bibitem{feifeipaper}
\bibinfo{author}{Qian, F.} \emph{et~al.}
\newblock \bibinfo{title}{Principles of appendage design in robots and animals
  determining terradynamic performance on flowable ground}.
\newblock \emph{\bibinfo{journal}{Bioinspiration and Biomimetics}}
  \textbf{\bibinfo{volume}{10}} (\bibinfo{year}{2015}).

\bibitem{li2013terradynamics}
\bibinfo{author}{Li, C.}, \bibinfo{author}{Zhang, T.} \&
  \bibinfo{author}{Goldman, D.~I.}
\newblock \bibinfo{title}{A terradynamics of legged locomotion on granular
  media}.
\newblock \emph{\bibinfo{journal}{Science}} \textbf{\bibinfo{volume}{339}},
  \bibinfo{pages}{1408--1412} (\bibinfo{year}{2013}).

\bibitem{zhang2013ground}
\bibinfo{author}{Zhang, T.} \emph{et~al.}
\newblock \bibinfo{title}{Ground fluidization promotes rapid running of a
  lightweight robot}.
\newblock \emph{\bibinfo{journal}{The International Journal of Robotics
  Research}} \textbf{\bibinfo{volume}{32}}, \bibinfo{pages}{859--869}
  (\bibinfo{year}{2013}).

\bibitem{li2012multi}
\bibinfo{author}{Li, C.}, \bibinfo{author}{Hsieh, S.~T.} \&
  \bibinfo{author}{Goldman, D.~I.}
\newblock \bibinfo{title}{Multi-functional foot use during running in the
  zebra-tailed lizard (callisaurus draconoides)}.
\newblock \emph{\bibinfo{journal}{The Journal of experimental biology}}
  \textbf{\bibinfo{volume}{215}}, \bibinfo{pages}{3293--3308}
  (\bibinfo{year}{2012}).

\bibitem{moritz2005human}
\bibinfo{author}{Moritz, C.~T.} \& \bibinfo{author}{Farley, C.~T.}
\newblock \bibinfo{title}{Human hopping on very soft elastic surfaces:
  implications for muscle pre-stretch and elastic energy storage in
  locomotion}.
\newblock \emph{\bibinfo{journal}{Journal of Experimental Biology}}
  \textbf{\bibinfo{volume}{208}}, \bibinfo{pages}{939--949}
  (\bibinfo{year}{2005}).

\bibitem{maladen2011mechanical}
\bibinfo{author}{Maladen, R.~D.}, \bibinfo{author}{Ding, Y.},
  \bibinfo{author}{Umbanhowar, P.~B.}, \bibinfo{author}{Kamor, A.} \&
  \bibinfo{author}{Goldman, D.~I.}
\newblock \bibinfo{title}{Mechanical models of sandfish locomotion reveal
  principles of high performance subsurface sand-swimming}.
\newblock \emph{\bibinfo{journal}{Journal of The Royal Society Interface}}
  \textbf{\bibinfo{volume}{8}}, \bibinfo{pages}{1332--1345}
  (\bibinfo{year}{2011}).

\bibitem{maladen2009undulatory}
\bibinfo{author}{Maladen, R.~D.}, \bibinfo{author}{Ding, Y.},
  \bibinfo{author}{Li, C.} \& \bibinfo{author}{Goldman, D.~I.}
\newblock \bibinfo{title}{Undulatory swimming in sand: subsurface locomotion of
  the sandfish lizard}.
\newblock \emph{\bibinfo{journal}{science}} \textbf{\bibinfo{volume}{325}},
  \bibinfo{pages}{314--318} (\bibinfo{year}{2009}).

\bibitem{Katsuragi2007}
\bibinfo{author}{Katsuragi, H.} \& \bibinfo{author}{Durian, D.~J.}
\newblock \bibinfo{title}{Unified force law for granular impact cratering}.
\newblock \emph{\bibinfo{journal}{Nature Physics}}
  \textbf{\bibinfo{volume}{3}}, \bibinfo{pages}{420--423}
  (\bibinfo{year}{2007}).
\newblock \urlprefix\url{http://www.nature.com/doifinder/10.1038/nphys583}.

\bibitem{tsimring2005impact}
\bibinfo{author}{Tsimring, L.} \& \bibinfo{author}{Volfson, D.}
\newblock \bibinfo{title}{Modeling of impact cratering in granular media}.
\newblock \emph{\bibinfo{journal}{Powders and grains}}
  \textbf{\bibinfo{volume}{2}}, \bibinfo{pages}{1215--1223}
  (\bibinfo{year}{2005}).

\bibitem{Umbanhowar2010}
\bibinfo{author}{Umbanhowar, P.} \& \bibinfo{author}{Goldman, D.}
\newblock \bibinfo{title}{Granular impact and the critical packing state}.
\newblock \emph{\bibinfo{journal}{Physical Review E}}
  \textbf{\bibinfo{volume}{82}}, \bibinfo{pages}{1--4} (\bibinfo{year}{2010}).
\newblock \urlprefix\url{http://link.aps.org/doi/10.1103/PhysRevE.82.010301}.

\bibitem{euler}
\bibinfo{author}{Euler, L.}
\newblock \bibinfo{title}{Neue grunds{\"a}tze der artillerie; reprinted in
  euler's opera omnia}.
\newblock \emph{\bibinfo{journal}{Druck und Verlag Von B.G. Teubner}}
  \textbf{\bibinfo{volume}{2}}, \bibinfo{pages}{1--409} (\bibinfo{year}{1922}).

\bibitem{poncelet}
\bibinfo{author}{Poncelet, J.~V.}
\newblock \emph{\bibinfo{title}{Cours de Mecanique Industrielle}}
  (\bibinfo{publisher}{Lithographie de Clouet, Paris}, \bibinfo{year}{1829}).

\bibitem{robins1972new}
\bibinfo{author}{Robins, B.} \& \bibinfo{author}{Curtis, W.}
\newblock \emph{\bibinfo{title}{New principles of gunnery}}
  (\bibinfo{publisher}{Richmond Publishing Company Limited},
  \bibinfo{year}{1972}).

\bibitem{backman1978mechanics}
\bibinfo{author}{Backman, M.~E.} \& \bibinfo{author}{Goldsmith, W.}
\newblock \bibinfo{title}{The mechanics of penetration of projectiles into
  targets}.
\newblock \emph{\bibinfo{journal}{International Journal of Engineering
  Science}} \textbf{\bibinfo{volume}{16}}, \bibinfo{pages}{1--99}
  (\bibinfo{year}{1978}).

\bibitem{allen1957dynamics}
\bibinfo{author}{Allen, W.~A.}, \bibinfo{author}{Mayfield, E.~B.} \&
  \bibinfo{author}{Morrison, H.~L.}
\newblock \bibinfo{title}{Dynamics of a projectile penetrating sand}.
\newblock \emph{\bibinfo{journal}{Journal of Applied Physics}}
  \textbf{\bibinfo{volume}{28}}, \bibinfo{pages}{370--376}
  (\bibinfo{year}{1957}).

\bibitem{forrestal1992penetration}
\bibinfo{author}{Forrestal, M.} \& \bibinfo{author}{Luk, V.}
\newblock \bibinfo{title}{Penetration into soil targets}.
\newblock \emph{\bibinfo{journal}{International Journal of Impact Engineering}}
  \textbf{\bibinfo{volume}{12}}, \bibinfo{pages}{427--444}
  (\bibinfo{year}{1992}).

\bibitem{pouliquen2009non}
\bibinfo{author}{Pouliquen, O.} \& \bibinfo{author}{Forterre, Y.}
\newblock \bibinfo{title}{A non-local rheology for dense granular flows}.
\newblock \emph{\bibinfo{journal}{Philosophical Transactions of the Royal
  Society of London A: Mathematical, Physical and Engineering Sciences}}
  \textbf{\bibinfo{volume}{367}}, \bibinfo{pages}{5091--5107}
  (\bibinfo{year}{2009}).

\bibitem{waitukaitis2012impact}
\bibinfo{author}{Waitukaitis, S.~R.} \& \bibinfo{author}{Jaeger, H.~M.}
\newblock \bibinfo{title}{Impact-activated solidification of dense suspensions
  via dynamic jamming fronts}.
\newblock \emph{\bibinfo{journal}{Nature}} \textbf{\bibinfo{volume}{487}},
  \bibinfo{pages}{205--209} (\bibinfo{year}{2012}).

\bibitem{Katsuragi2013}
\bibinfo{author}{Katsuragi, H.} \& \bibinfo{author}{Durian, D.~J.}
\newblock \bibinfo{title}{Drag force scaling for penetration into granular
  media}.
\newblock \emph{\bibinfo{journal}{Physical Review E}}
  \textbf{\bibinfo{volume}{87}}, \bibinfo{pages}{052208}
  (\bibinfo{year}{2013}).
\newblock \urlprefix\url{http://link.aps.org/doi/10.1103/PhysRevE.87.052208}.

\bibitem{brennen1982review}
\bibinfo{author}{Brennen, C.}
\newblock \bibinfo{title}{A review of added mass and fluid inertial forces.}
\newblock \bibinfo{type}{Tech. Rep.}, \bibinfo{institution}{DTIC Document}
  (\bibinfo{year}{1982}).

\bibitem{pandy1990optimal}
\bibinfo{author}{Pandy, M.~G.}, \bibinfo{author}{Zajac, F.~E.},
  \bibinfo{author}{Sim, E.} \& \bibinfo{author}{Levine, W.~S.}
\newblock \bibinfo{title}{An optimal control model for maximum-height human
  jumping}.
\newblock \emph{\bibinfo{journal}{Journal of biomechanics}}
  \textbf{\bibinfo{volume}{23}}, \bibinfo{pages}{1185--1198}
  (\bibinfo{year}{1990}).

\bibitem{zajac1993muscle}
\bibinfo{author}{Zajac, F.~E.}
\newblock \bibinfo{title}{Muscle coordination of movement: a perspective}.
\newblock \emph{\bibinfo{journal}{Journal of Biomechanics}}
  \textbf{\bibinfo{volume}{26}}, \bibinfo{pages}{109--124}
  (\bibinfo{year}{1993}).

\bibitem{aguilar2012lift}
\bibinfo{author}{Aguilar, J.}, \bibinfo{author}{Lesov, A.},
  \bibinfo{author}{Wiesenfeld, K.} \& \bibinfo{author}{Goldman, D.~I.}
\newblock \bibinfo{title}{Lift-off dynamics in a simple jumping robot}.
\newblock \emph{\bibinfo{journal}{Physical review letters}}
  \textbf{\bibinfo{volume}{109}}, \bibinfo{pages}{174301}
  (\bibinfo{year}{2012}).

\bibitem{locrobophys}
\bibinfo{author}{Aguilar, J.} \emph{et~al.}
\newblock \bibinfo{title}{A review on locomotion robophysics: the study of
  movement at the intersection of robotics, soft matter and dynamical systems,
  in press}.
\newblock \emph{\bibinfo{journal}{Reports on Progress in Physics}}
  (\bibinfo{year}{2015}).

\bibitem{gravish2014force}
\bibinfo{author}{Gravish, N.}, \bibinfo{author}{Umbanhowar, P.~B.} \&
  \bibinfo{author}{Goldman, D.~I.}
\newblock \bibinfo{title}{Force and flow at the onset of drag in plowed
  granular media}.
\newblock \emph{\bibinfo{journal}{Physical Review E}}
  \textbf{\bibinfo{volume}{89}}, \bibinfo{pages}{042202}
  (\bibinfo{year}{2014}).

\bibitem{tapia2013effect}
\bibinfo{author}{Tapia, F.}, \bibinfo{author}{Esp{\'\i}ndola, D.},
  \bibinfo{author}{Hamm, E.} \& \bibinfo{author}{Melo, F.}
\newblock \bibinfo{title}{Effect of packing fraction on shear band formation in
  a granular material forced by a penetrometer}.
\newblock \emph{\bibinfo{journal}{Physical Review E}}
  \textbf{\bibinfo{volume}{87}}, \bibinfo{pages}{014201}
  (\bibinfo{year}{2013}).

\bibitem{stone2004stress}
\bibinfo{author}{Stone, M.~B.} \emph{et~al.}
\newblock \bibinfo{title}{Stress propagation: Getting to the bottom of a
  granular medium}.
\newblock \emph{\bibinfo{journal}{Nature}} \textbf{\bibinfo{volume}{427}},
  \bibinfo{pages}{503--504} (\bibinfo{year}{2004}).

\bibitem{stone2004local}
\bibinfo{author}{Stone, M.} \emph{et~al.}
\newblock \bibinfo{title}{Local jamming via penetration of a granular medium}.
\newblock \emph{\bibinfo{journal}{Physical Review E}}
  \textbf{\bibinfo{volume}{70}}, \bibinfo{pages}{041301}
  (\bibinfo{year}{2004}).

\bibitem{le2014emergence}
\bibinfo{author}{Le~Bouil, A.}, \bibinfo{author}{Amon, A.},
  \bibinfo{author}{McNamara, S.} \& \bibinfo{author}{Crassous, J.}
\newblock \bibinfo{title}{Emergence of cooperativity in plasticity of soft
  glassy materials}.
\newblock \emph{\bibinfo{journal}{Physical review letters}}
  \textbf{\bibinfo{volume}{112}}, \bibinfo{pages}{246001}
  (\bibinfo{year}{2014}).

\bibitem{glasheen1996hydrodynamic}
\bibinfo{author}{Glasheen, J.} \& \bibinfo{author}{McMahon, T.}
\newblock \bibinfo{title}{A hydrodynamic model of locomotion in the basilisk
  lizard}.
\newblock \emph{\bibinfo{journal}{Nature}} \textbf{\bibinfo{volume}{380}},
  \bibinfo{pages}{340--341} (\bibinfo{year}{1996}).

\bibitem{richardson1948impact}
\bibinfo{author}{Richardson, E.}
\newblock \bibinfo{title}{The impact of a solid on a liquid surface}.
\newblock \emph{\bibinfo{journal}{Proceedings of the Physical Society}}
  \textbf{\bibinfo{volume}{61}}, \bibinfo{pages}{352--367}
  (\bibinfo{year}{1948}).

\bibitem{wagner1932phenomena}
\bibinfo{author}{Wagner, H.}
\newblock \bibinfo{title}{Phenomena associated with impacts and sliding on
  liquid surfaces}.
\newblock \emph{\bibinfo{journal}{Z. Angew. Math. Mech}}
  \textbf{\bibinfo{volume}{12}}, \bibinfo{pages}{193--215}
  (\bibinfo{year}{1932}).

\bibitem{sakakibara2004stereo}
\bibinfo{author}{Sakakibara, J.}, \bibinfo{author}{Nakagawa, M.} \&
  \bibinfo{author}{Yoshida, M.}
\newblock \bibinfo{title}{Stereo-piv study of flow around a maneuvering fish}.
\newblock \emph{\bibinfo{journal}{Experiments in fluids}}
  \textbf{\bibinfo{volume}{36}}, \bibinfo{pages}{282--293}
  (\bibinfo{year}{2004}).

\bibitem{clark2013granular}
\bibinfo{author}{Clark, A.~H.} \& \bibinfo{author}{Behringer, R.~P.}
\newblock \bibinfo{title}{Granular impact model as an energy-depth relation}.
\newblock \emph{\bibinfo{journal}{EPL (Europhysics Letters)}}
  \textbf{\bibinfo{volume}{101}}, \bibinfo{pages}{64001}
  (\bibinfo{year}{2013}).

\bibitem{zhang2001zeno}
\bibinfo{author}{Zhang, J.}, \bibinfo{author}{Johansson, K.~H.},
  \bibinfo{author}{Lygeros, J.} \& \bibinfo{author}{Sastry, S.}
\newblock \bibinfo{title}{Zeno hybrid systems}.
\newblock \emph{\bibinfo{journal}{International Journal of Robust and Nonlinear
  Control}} \textbf{\bibinfo{volume}{11}}, \bibinfo{pages}{435--451}
  (\bibinfo{year}{2001}).

\bibitem{francois2013geometrical}
\bibinfo{author}{Francois, N.}, \bibinfo{author}{Saadatfar, M.},
  \bibinfo{author}{Cruikshank, R.} \& \bibinfo{author}{Sheppard, A.}
\newblock \bibinfo{title}{Geometrical frustration in amorphous and partially
  crystallized packings of spheres}.
\newblock \emph{\bibinfo{journal}{Physical review letters}}
  \textbf{\bibinfo{volume}{111}}, \bibinfo{pages}{148001}
  (\bibinfo{year}{2013}).

\bibitem{daniels2005hysteresis}
\bibinfo{author}{Daniels, K.~E.} \& \bibinfo{author}{Behringer, R.~P.}
\newblock \bibinfo{title}{Hysteresis and competition between disorder and
  crystallization in sheared and vibrated granular flow}.
\newblock \emph{\bibinfo{journal}{Physical review letters}}
  \textbf{\bibinfo{volume}{94}}, \bibinfo{pages}{168001}
  (\bibinfo{year}{2005}).

\end{thebibliography}

\clearpage
\begin{figure}[t]
\begin{large}
\textbf{Figures}
\end{large}

\begin{centering}
\includegraphics{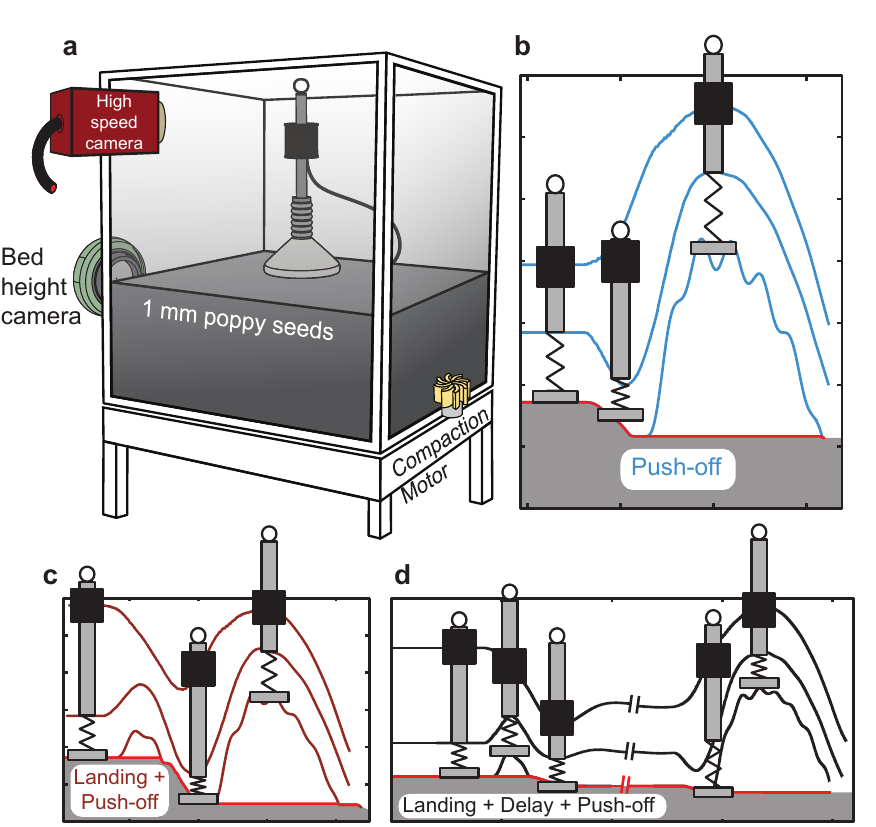}
\caption{\textbf{An actively and passively self-deforming robot jumping on granular media.} a, $\sim$1 mm diameter poppy seeds fill a 56 cm x 56 cm area fluidized bed to a height of $\sim 25$ cm. Volume fraction is controlled with a compaction motor and air flow, and the robot is constrained via an air bearing (not shown) to jump vertically. Simulated (using an experimentally validated numerical integration of robot equations of motion and equation (4) for granular forces, see methods) time series illustrations (of foot, rod and motor) show jumping trajectories for a push-off intrusion, or single jump (b), landing and push-off, or stutter jump (c), and landing, delay and push-off, or delayed stutter jump (d). Robot size scaled down by $\sim$ 4x for illustrative purposes.}
\end{centering}
\end{figure}

\afterpage{
\thispagestyle{empty}
\pagestyle{empty} 
\begin{figure}[ht]
\begin{centering}
\includegraphics{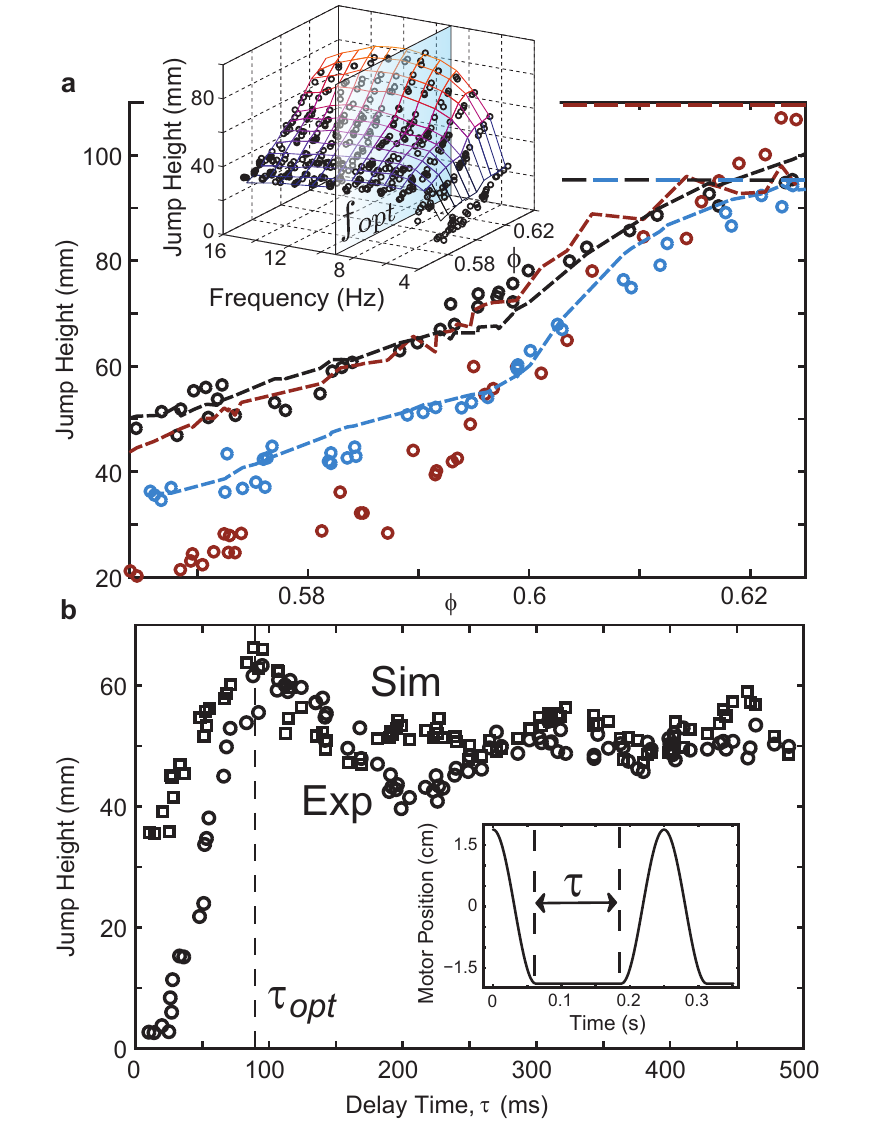}
\caption{\textbf{Jump heights for various self-deformations.} a, Experimental jump heights at optimal forcing frequency ($f_{opt}$ determined according to highest jump at high $\phi$) (circles) vs. $\phi$ compared with 1D simulation (dashed lines) results using the traditional granular force relation, equation (1) (and the 2-resistance re-intrusion relation for $F_p(z)$), for single jumps (blue), stutter jumps (maroon), and delayed stutter jumps (black). Hard ground jump heights are indicated by horizontal dashed lines. Each jump type is produced with a sine wave at optimal frequency determined from a larger sweep of forcing frequencies (single jump data in inset). b, Simulation (squares) and experimental (circles) heights of delayed stutters in loose poppy ($\phi=0.57$) agree for delay times, $\tau\geq\tau_{opt}$.}
\end{centering}
\end{figure}
\clearpage
}

\afterpage{
\begin{figure}[ht]
\begin{centering}
\includegraphics[width=.7\textwidth]{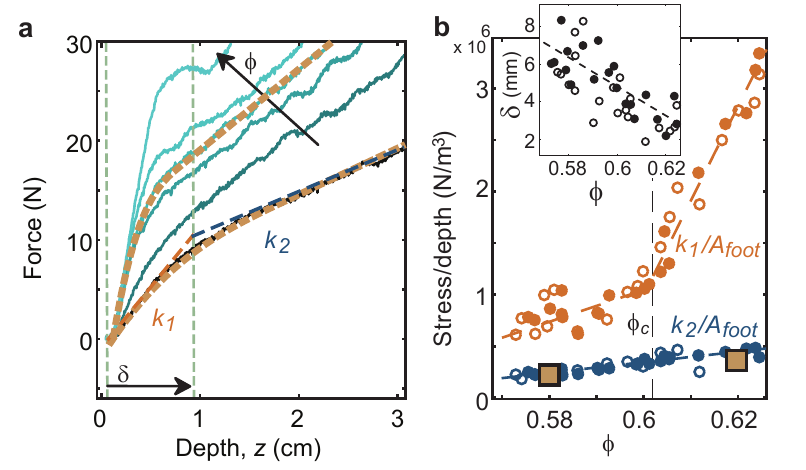}
\caption{\textbf{Empirical measurements of force vs. intrusion depth.} A 5.1 cm diameter (a, b solid circles) and 7.6 cm diameter (b open circles) flat foot were tested for increasing $\phi$ (black to light blue in a). Transition depths, $\delta$, between low and high penetration resistance vs. $\phi$ are displayed in inset in (b). Empirical estimates of $k_1/A_{foot}$, RFT measurements for angled intrusions\cite{li2013terradynamics} and a model of granular cone jamming (Fig. 5a) were combined to predict force vs. depth at $\phi=0.58$ and $\phi=0.62$ (a, brown dashed curves). Stress vs. depth for fully developed cones accurately predicted the $k_2/A_{foot}$ penetration resistance at low and high $\phi$ (b, brown squares), where $A_{foot}$ is the foot surface area.}
\end{centering}
\end{figure}
\clearpage
}

\afterpage{
\begin{figure*}[ht]
\begin{centering}
\includegraphics{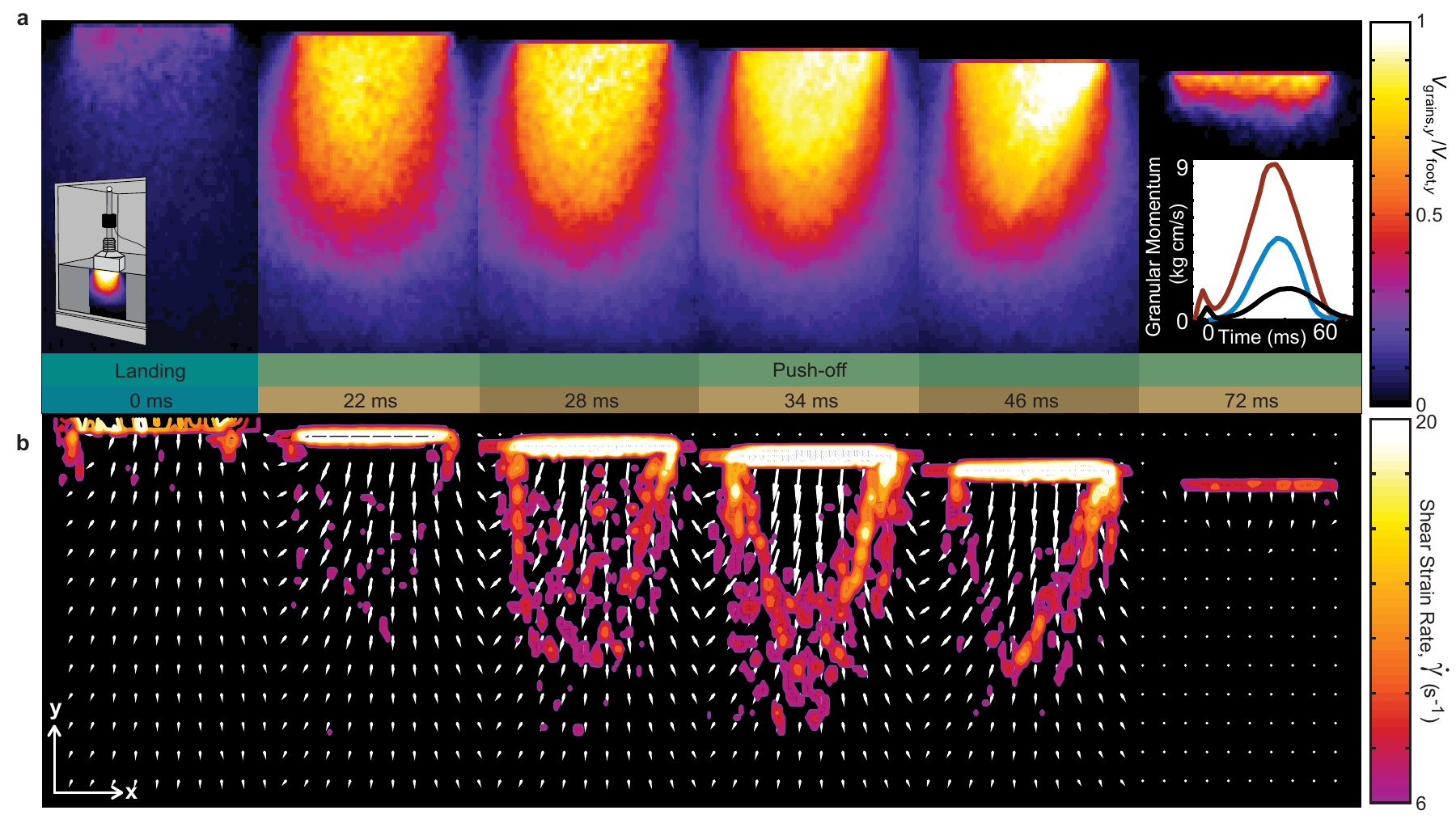}
\caption{\textbf{Particle Image Velocimetry (PIV) measurement of granular flow kinematics.} a, Frame sequence of the downward velocity field, normalized by foot speed, taken during the landing and push-off phase of a stutter jump at $\phi=0.57$ (left inset, diagram of setup). Right inset, Time sequence of granular momentum calculated from PIV for single (blue), stutter (maroon) and delayed stutter (black) jumps at $\phi=0.57$. b, PIV vector field of same snapshots superimposed by shear bands derived from the shear strain field according to equation (2). Shears bands illustrate how a cone of jammed grains rapidly emerge below the foot and wedge through surrounding material.}
\end{centering}
\end{figure*}
\clearpage
}

\afterpage{
\begin{figure}[h]
\begin{centering}
\includegraphics[width=.7\textwidth]{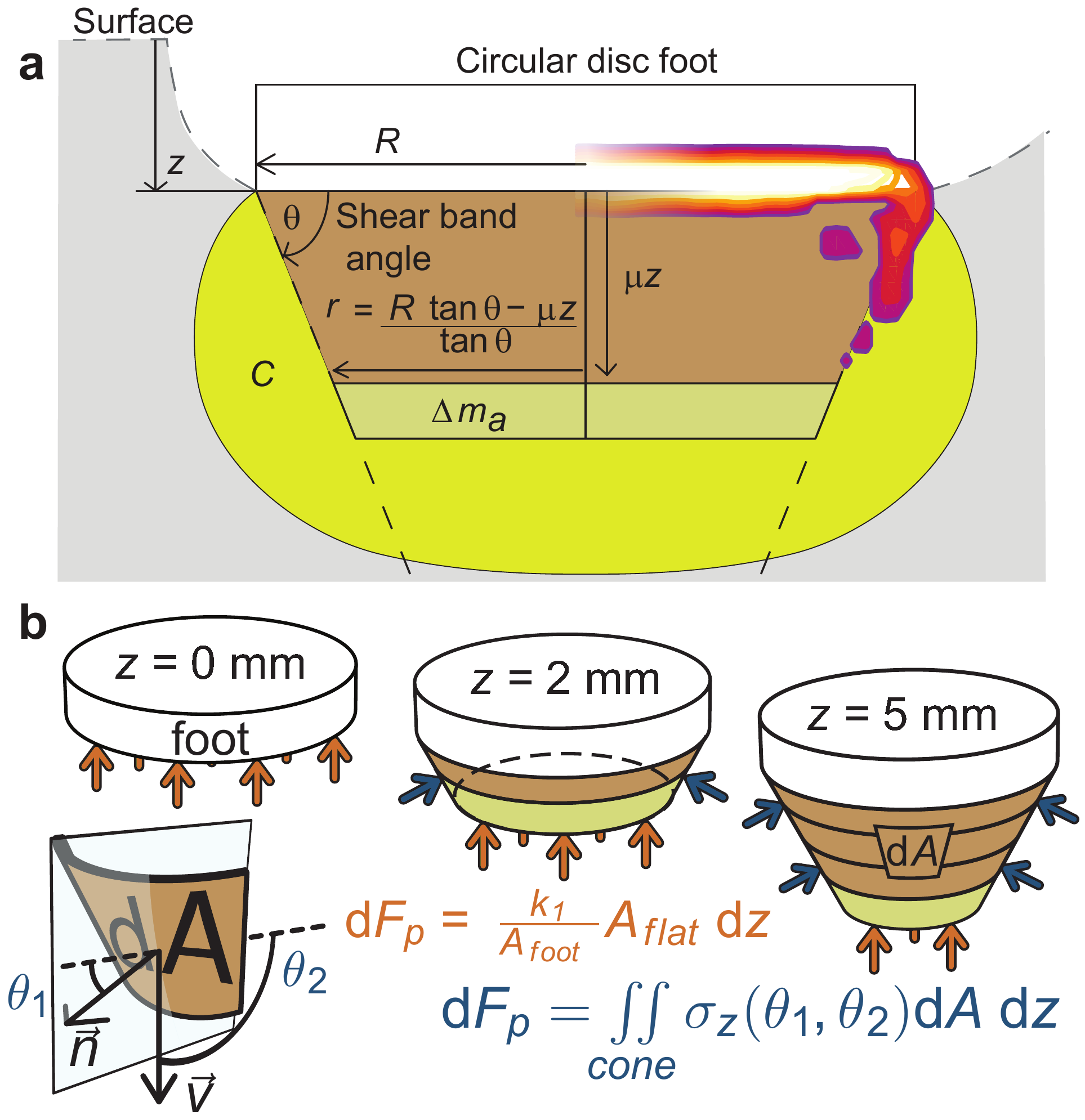}
\caption{\textbf{Quasistatic and inertial properties of a jamming granular cone.} a, A geometric model of granular cone evolution vs. intrusion depth. An added mass model of this cone takes into account a solidified conical core (brown) as well as extra virtual mass, $C$, from slower moving grains surrounding the cone (yellow). The conical angle, $\theta$, is estimated from the angle of shear bands from PIV (superimposed). b, Quasistatic force contributions, $F_p(z)$ at different stages of cone evolution. Flat surfaces (orange) were estimated with the empirical $k_1$ penetration resistance. Angled conical surface forces (blue) were calculated using the RFT stress model\cite{li2013terradynamics}, for vertical stress, $\sigma_z$, on a differential surface element, $\delta A$, at an angle, $\phi_1 = 60^\circ$, for the orientation normal vector, $\vec{n}$, and an angle, $\phi_2 = 90^\circ$, for the velocity vector, $\vec{v}$.}
\end{centering}
\end{figure}
\clearpage
}

\afterpage{
\begin{figure}[h]
\begin{centering}
\includegraphics[width = .45\hsize]{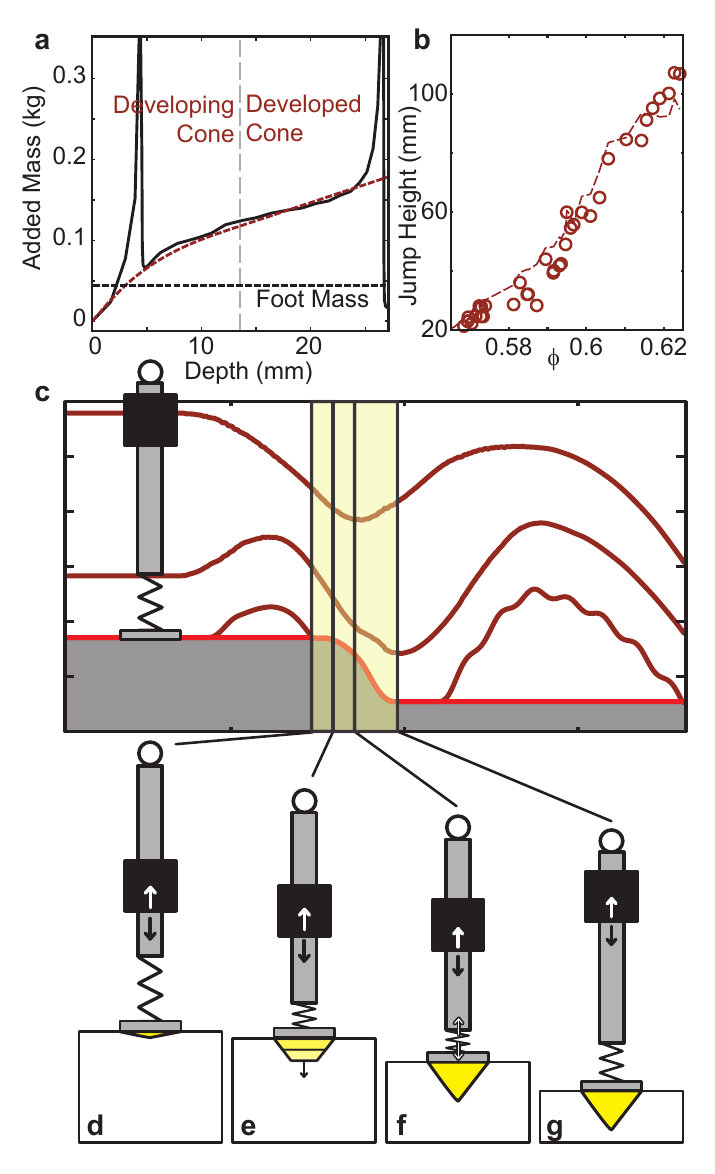}
\caption{\textbf{Simulation of coupled added mass and robot jumping dynamics.} a, Added mass vs. depth calculation from PIV (black, solid) and a saturating cone equation (brown, dashed, derived from the geometric model in Fig. 5a) for a stutter jump at $\phi=0.57$. b, A simulation (dashed line) of stutter jump heights vs. $\phi$ using equation (4) for granular forces improves agreement with experiment (circles) at low $\phi$. c, Time trajectories of the motor, rod and foot positions using 1-D simulations at $\phi=0.57$. d-g, Snapshots of robot during landing and push-off illustrate the interplay of granular forces on stutter jump dynamics, from (d) initial foot landing, to (e) rapid added mass recruitment, to (f) spring decompression (white arrows) and a fully developed cone, to (g) granular jamming. Relative positions of robot elements were taken from 1-D simulation. Arrows on rod and motor indicate the rod being pushed down relative to the motor. Yellow added mass regions are illustrated based on the experimental PIV observations; such observations inspired the model of added mass included in the 1-D simulation. Robot scaled down by $\sim$ 4x for illustrative purposes.}
\end{centering}
\end{figure}
\clearpage
}

\clearpage

\begin{methods}

\textbf{Robot Jumper.} We performed systematic experiments on a robo-physics style jumping robot in a fluidizable bed of $\sim$1 mm poppy seeds. The apparatus (Fig. 1a) was fully automated, allowing for simultaneous robot control and data acquisition while sequentially exploring a parameter space. The robotic jumper was adapted from previous hard ground experiments\cite{aguilar2012lift}, consisting of a Dunkermotoren STA-1104 linear actuator connected to the carriage of an air bearing that allowed for nearly frictionless motion constrained in the vertical direction. The actuator-carriage unit had a mass of 1.125 kg and comprised the majority of the robot mass. The actuator applied a force proportional to the supplied current to a 0.125 kg thrust rod. The actuator maintained a commanded position relative to the rod by supplying the appropriate voltage according to a feedback control protocol. The bottom of the rod was connected to a spring with stiffness $k_s$ = 3300 N/m. The bottom of the spring was connected to a 7.6 cm diameter flat disc foot. To produce various jumping movements, the motor followed a one-cycle sine wave positional trajectory with an amplitude $A = 1.875$ cm. During jumping, the centroid position of a 9.5 mm white plastic ball fixed to the thrust rod was captured by a 200 fps camera to track rod position, and the jump height was calculated as the maximal rod position minus the initial rod position at rest. For the delayed stutter jump with the maximal wait time (0.75 sec), vibrational transients were eliminated by temporarily lowering the proportional positional feedback gain in the linear motor, producing an amplified damping effect in the spring vibration.

\textbf{Fluidized Bed.} The entire jumping/air bearing assembly was placed inside a bed of granular media. To set the compaction of the granular media, the substrate was air fluidized by a 5 hp blower with variable voltage flow control that sends air flow to the bottom of the bed through a Porex flow diffuser. This fluidization process reset the state of media from any previous disturbances and produced a loose-packed state with volume fraction, $\phi \approx 0.57$. Producing higher compactions consisted of modulating air-flow rate below onset of fluidization to produce air pulses while simultaneously activating a shaker motor that vibrated the bed. Volume fractions, measured with a camera that captures bed height, ranged from 0.57 to 0.62. A separate linear motor lifted the jumper during this granular preparation process between jumping experiments.

\textbf{1D Jumping Model.} We numerically integrated a Simulink (Matlab) model of a self-deforming actuator (comprised of a linear motor and thrust rod) in series with a spring and foot jumping on granular media according to the following equations of motion: $m_{m}\ddot{x}_{m} = -m_{m}g+F_{m}$, $m_{r}\ddot{x}_{r} = -m_{r}g-F_{m}+F_s$, $m_{f}\ddot{x}_{f} = -m_{f}g-F_s+F_{GM}$, the subscripts, \textit{m}, \textit{r} and \textit{f} corresponding to motor, rod and foot quantities, respectively. The rod and motor equations were combined as: $M\ddot{x}_{r} = -Mg+F_s + m_{m}\ddot{X}_{m}$, where $M=m_{m}+m_{r}$, and the $\ddot{X}_m=\ddot{x}_{r}-\ddot{x}_{m}$. To compare directly with experiment, we used experimental encoder positions from each experimental jump performed to obtain the $\ddot{X}_m$ command for simulation. The granular force, $F_{GM}$, followed the various relations discussed in this Article, and the spring force, $F_s$, followed Hooke's law for the spring between the rod and foot. Another property of the $F_{GM}$ was its hybrid dynamic dependence on the discrete transition of the foot between the ground and air phases (i.e. $F_{GM}=0$ during aerial phase). The ground position changed as the foot intruded and was set to the foot's position while the foot was grounded. During the robot's aerial phase, the ground maintained the last foot position before transition to the aerial phase.

A challenge of numerically integrating a 1D damped bouncing system even on hard ground is mitigating so-called ``Zeno'' effects, in which the number of bounces approaches infinity in finite time\cite{zhang2001zeno}. This leads to significant simulation errors which scale with time-step size in detecting the transition between ground and aerial phases. To reduce such inaccuracies, we used Matlab's ODE45 integrator, which has a variable time-step that is adjusted according the current system stiffness, thus accurately detecting hybrid transitions. Moreover, on granular media, the ground position changes during the grounded phase, which can cause perpetual Zeno-like behavior unless proper conditions are established for determining the transition from the ground to aerial phase.

A naive approach is to state that the foot becomes aerial when the total force on the foot causes the foot to accelerate from a negative to positive velocity. However, this caused constant Zeno-like switching between the ground and air phase and was only accurate for extremely small integration time-steps. An understanding of the nature of the different forces is required to obtain the proper transition conditions. Ground reaction forces, or $F_{GM}$ in this case, are merely capable of resisting downward motion, and not producing propulsive upward motion. Thus, even if $\dot{x}_f=0$ and $\ddot{x}_f>0$, the foot remains grounded and foot speed remains zero at the next time step if the spring is still compressed and pushing down on the foot ($-F_s-m_fg<0$). To become aerial, the foot must be pulled off the ground by the spring ($-F_s-m_fg>0$), rather than pushed off by the ground. This condition eliminated Zeno while achieving output results identical to the small time-step approach, yielding faster simulation.

\textbf{Fitting procedure for granular force models in 1D simulation.} When using equation (1) in the Article, we fit simulation jump heights to experiment with a combination of empirical measurements and systematic parameter fitting. $F_p(z)$ was primarily determined via slow-velocity force vs. depth measurements. In our two-resistance formulation, $k_1$, $k_2$, and $\delta$ were determined with respect to $\phi$ by systematically performing intrusions at various values of $\phi$. We then determined a depth-independent $\alpha$ at each $\phi$ by systematically varying $\alpha$ while simulating single jumps at each $\phi$ and compared simulation jump heights vs. forcing frequency to experiment. Jump heights were most sensitive to $\alpha$ for high forcing frequencies, as expected since higher frequencies induced higher intrusion speeds. Fits of $\alpha$ vs. $\phi$ revealed a similar scaling with $\phi$ as $k_1$: a higher $d\alpha/d\phi$ was observed for $\phi>\phi_c$. This fit of $\alpha$ with a two-resistance empirical $F_p(z)$ yielded good agreement between simulation and experiment single jumps.

However, a comparison of simulation and experiment of delayed stutter jumps and stutter jumps revealed poor agreement for the $\alpha$ fits and $F_p(z)$. To determine if modification to $F_p(z)$ was needed due to reintrusions, we performed reintrusion measurements by intruding the thrust rod at slow velocities to a certain depth, extracting, and then reintruding. Upon reintrusion, we observed a sharp rise in force to a peak that was higher than the expected force according to the original two-resistance force relation. This led to a new formulation of $F_p(z)$ according to the observed reintrusion force behavior that had a force overshoot proportional to the depth of reintrusion. We performed a fitting process similar to $\alpha$ fitting to attain the correct reintrusion parameters, leading to simulation agreement with both single jumps as well as delayed stutter jumps with delay times $\tau\geq\tau_{opt}$. For stutter jumps with delay $\tau<\tau_{opt}$, simulation still did not agree with experiment.

This led to the use of a granular force model that incorporates more complex inertial effects (equation (4)), which includes an added mass force and a depth-dependant $\alpha$ for inertial drag. The added mass force directly multiplies the foot's acceleration with the added mass, $m_a$ which is determined by equation (6). We varied $\mu$ and $C$ in equation (6) to match empirical added mass vs. depth measurements for all depths except when the foot speed approached zero, which caused a singularity in the added mass measurement. However, since this rapid increase in added mass only occurs during slow velocities, its effect on jumping dynamics is negligible; this was confirmed by simulations incorporating the singularity. For the inertial drag force, $\alpha(z) = b\frac{dm_a}{dz}$, we set the scaling coefficient, $b$, for each $\phi$ such that there was agreement between experiment and simulation for all jumps. Interestingly, $b$ tended to increase with $\phi$ in a similar qualitative trend to $k_1$.

\textbf{Static Intrusion Force Measurements.} To characterize the static penetration force, $F_p$, we repurposed the robot's motor for intrusion force measurements. With the motor clamped securely to the bed, the rod was connected directly to the foot and slowly forced at constant speed into poppy seeds at various $\phi$. Force and depth measurements were attained from motor current and encoder position, respectively.  We used both 5.1 cm and 7.6 cm diameter flat feet and found that $F_p$ scaled proportionally with foot surface area.

\textbf{PIV Experiment.} We moved the robot from the center of the granular bed to the clear Acrylic side wall, and, using a foot with a flat side, we had the robot perform all 3 jump strategies for a sparse sweep of volume fractions and recorded high speed video (500 fps AOS camera at 1280 x 1024 resolution) of the sidewall grain flow. Jump heights did not deviate significantly from jump heights at the center of the bed (away from wall effects). For PIV analysis, no tracer particles were necessary, since enhancing local contrast in poppy seed images provided a sufficiently large and well mixed distribution of grey-scale intensities among grains.

\end{methods}

\clearpage

\begin{large}
\noindent{\textbf{Supplementary Information for ``Robophysical study of jumping dynamics on granular media''}}
\end{large}

{\em \textbf{Supplementary I: 2D DEM Simulations}} -- 

To illustrate the geometric evolution of the cone jamming phenomenon, we developed a simple discrete element model (DEM) of 2D disks interacting. Grain-grain interactions consisted of a simple Hooke's law spring repulsion, zero friction, and viscous damper repulsion, which allowed for inelastic collisions. While we instituted a simplification of zero grain-grain friction, we introduced individual grain friction to prevent excessive inertial motion (which would resemble a billiards simulation), allowing for a more geometric analysis of cone evolution. A flat object was approximated by ``intruder grains'' that move with a prescribed trajectory and apply similar grain-grain interaction forces on other ``free'' grains. Accelerations were integrated using an explicit forward Euler integration scheme.

To compare with experiment, we simulated all three jump types (single, stutter and delayed stutter) by prescribing the intruder trajectory to be the trajectories of the foot tracked during the PIV experiments. For momentum and added mass calculations, bidisperse particles were used to avoid crystallization. Using a 2D grid of interpolated grain velocities, we were able to calculate added mass and momentum identically to PIV measurements. Simulation values for momentum and added mass were about twice those of PIV (Extended Data Fig. 2b). While properties such as grain size, density and foot size were chosen to resemble experimental values, there are numerous differences between the DEM simulation and PIV experiments that are likely contributing to the numerical discrepancies, from the difference in shape (Extended Data Fig. 2c) to the lack of gravity, boundaries or shearing grain-grain friction in simulation to the fact that the simulation is only considers 2D dynamics. However, regardless of these differences, even this simple billiards-like simulation was able to capture many of the qualitative observations in experiment, from the relative scaling of grain momentum between different jump types to the saturating added mass.

Additionally, the simulation revealed a similar triangular jamming front as that found with PIV experiment (Extended Data Fig. 2a) and shear bands observed in other compression experiments\cite{le2014emergence}. Using monodisperse grains produced crystallizing (a feature of monodisperse spherical particles~\cite{francois2013geometrical,daniels2005hysteresis}) triangles resulting from $60\degree$ shear bands. Similarly forming triangles emerged without crystallizing lattices using both bidisperse and polydisperse particles. A simple 2D DEM was able to give a clear picture of how grains jam with the foot through local crystalizations which shear sides as each layer of grains is deceased in length by one grain (Extended Data Fig. 2a).

\clearpage
\begin{figure}[t]
\begin{large}
\textbf{Extended Data Figures}
\end{large}

\begin{centering}
\includegraphics[width=.99\textwidth]{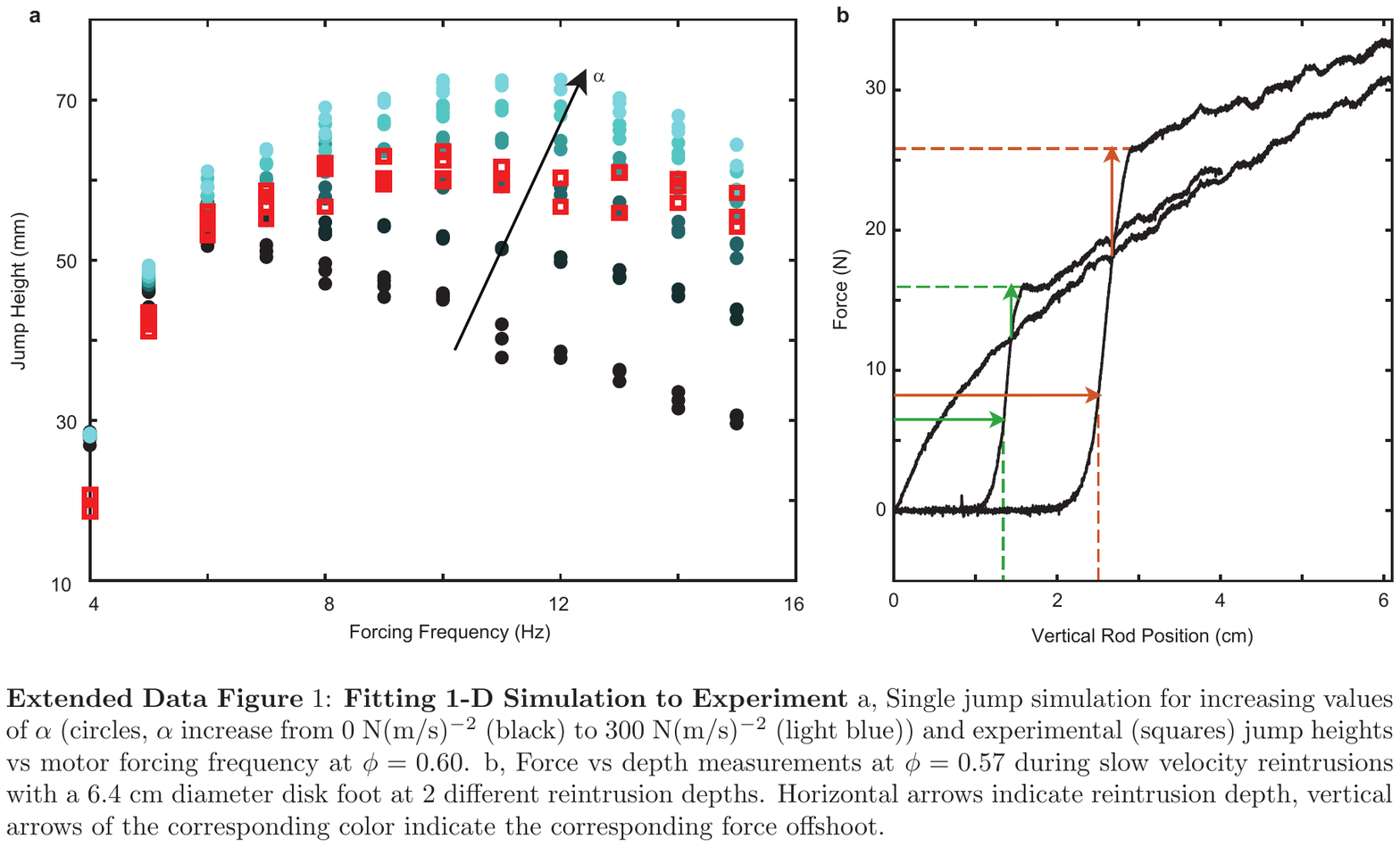}
\end{centering}
\end{figure}

\afterpage{
\begin{figure}[ht]
\begin{centering}
\includegraphics[width=.99\textwidth]{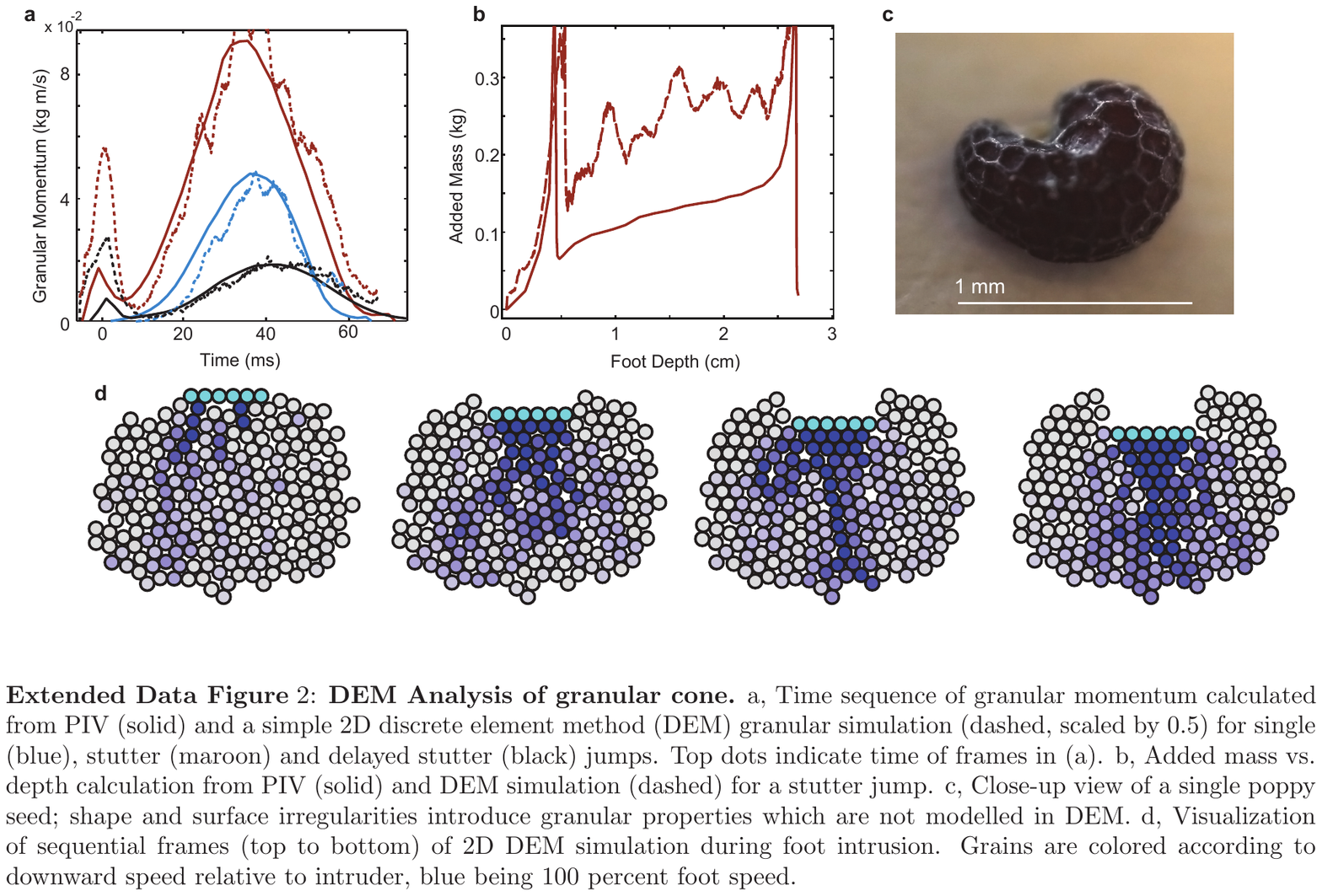}
\end{centering}
\end{figure}
\clearpage
}

\afterpage{
\begin{figure}[ht]
\begin{centering}
\includegraphics[width=.99\textwidth]{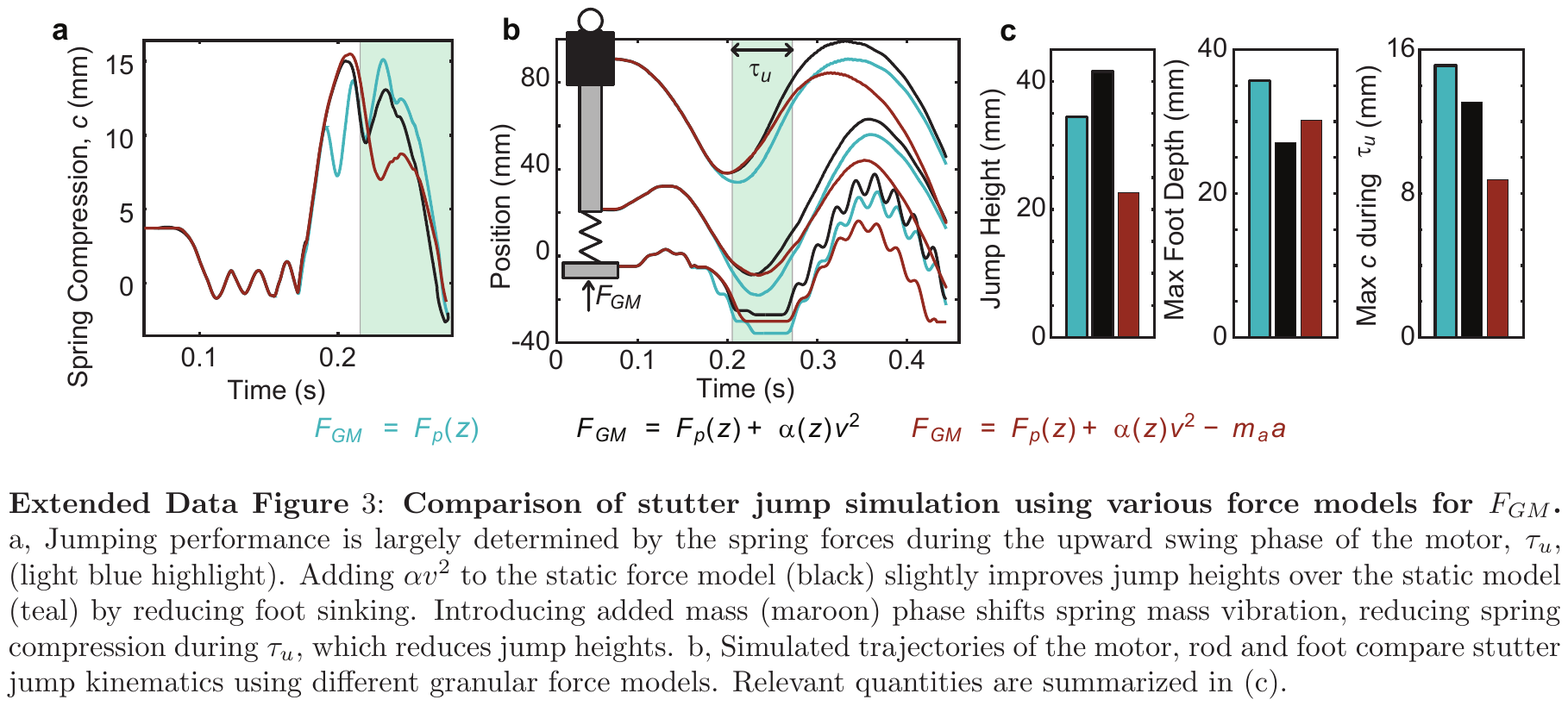}
\end{centering}
\end{figure}
\clearpage
}

\begin{large}
\noindent{\textbf{Supplementary Movies}}
\end{large}

\textbf{S1: High speed video of side wall grain flow during jumping.} -- Video sequentially displays poppy seed grain flow in loose ($\phi$=0.57) and close ($\phi$=0.62) packed preparations while the foot intrudes during single jumps, stutter jumps, and delayed stutter jumps. The delayed stutter jump has frames omitted between landing and push-off to speed up video.

\textbf{S2: PIV analysis of grains during stutter jump in loose packed ($\phi$=0.57) poppy seeds.} Simultaneous visualization of particle image velocimetry and jumping robot animated from video-tracked position of the thrust rod, relative encoder position of the motor, and video-tracked position of foot (tracked in same video as 500 fps sidewall grain flow video used for PIV). Stutter jump is first displayed with arrow PIV velocity field and color field of granular speed, and then the stutter jump is displayed with max shear strain rate field calculated from PIV (shear bands highlight the edges of the cone of grains moving with foot).
\end{document}